\documentclass[iop]{emulateapj}

\usepackage{natbib}
\usepackage{graphicx}
\usepackage{multirow}

\def\sun{\ifmmode\odot\else$\odot$\fi}

\newcommand{\cv}{$f_{\rm 2}$ }
\newcommand{\ncv}{$f_{\rm 2}$}
\newcommand{\mic}{$\mu$m}                      %

\defcitealias{mateos16}{M16}

\shorttitle{Survival of the obscuring torus in QSOs}\shortauthors{Mateos et al.}
\shortauthors{Mateos et al.}

\begin{document}

\title{Survival of the obscuring torus in the most powerful active galactic nuclei}
\author{
  S. Mateos\altaffilmark{1},
  F. J. Carrera\altaffilmark{1},
  X. Barcons\altaffilmark{1}, 
  A. Alonso-Herrero\altaffilmark{2},
  A. Hern\'an-Caballero\altaffilmark{3},   
  M. Page\altaffilmark{4},
  C. Ramos Almeida\altaffilmark{5,6}, 
  A. Caccianiga\altaffilmark{7},
  T. Miyaji\altaffilmark{8},
  A. Blain\altaffilmark{9}
}

\altaffiltext{1}{Instituto de F\'{\i}sica de
  Cantabria, CSIC-UC, 39005 Santander, Spain;
  E-mail: mateos@ifca.unican.es} 
\altaffiltext{2}{ Centro de Astrobiolog\'ia (CAB, CSIC-INTA), ESAC
  Campus, E-28692 Villanueva de la Ca\~nada, Madrid, Spain}
\altaffiltext{3}{ Departamento de Astrof\'isica, Facultad de
  CC. F\'isicas, Universidad Complutense de Madrid, 28040 Madrid,
  Spain}
\altaffiltext{4}{Mullard Space Science Laboratory, University College
  London, Holmbury St Mary, Dorking, Surrey RH5 6NT, UK}
\altaffiltext{5}{Instituto de Astrof\'isica de Canarias (IAC),
  E-38205 La Laguna, Tenerife, Spain}
\altaffiltext{6}{Departamento de Astrof\'isica, Universidad de la
  Laguna (ULL), E-38206 La Laguna, Tenerife, Spain}
\altaffiltext{7}{INAF - Osservatorio Astronomico di Brera, via Brera
  28, I-20121 Milan, Italy}
\altaffiltext{8}{Instituto de Astronom\'ia sede Ensenada, Universidad
  Nacional Aut\'onoma de M\'exico, Km. 103, Carret. Tijuana-Ensenada,
  Ensenada, BC 22860, M\'exico}
\altaffiltext{9}{Department of Physics and Astronomy, University of
  Leicester, Leicester, LE1 7RH, UK}

\begin{abstract}
  Dedicated searches generally find a decreasing fraction of obscured
  Active Galactic Nuclei (AGN) with increasing AGN luminosity. This
  has often been interpreted as evidence for a decrease of the
  covering factor of the AGN torus with increasing luminosity, the
  so-called receding torus models. Using a complete flux-limited X-ray
  selected sample of 199 AGN, from the Bright Ultra-hard XMM-Newton
  Survey, we determine the intrinsic fraction of optical type-2 AGN at
  0.05$\leq$$z$$\leq$1 as a function of rest-frame 2-10 keV X-ray
  luminosity from ${\rm 10^{42}}$ to ${\rm 10^{45}\,erg\,s^{-1}}$. We
  use the distributions of covering factors of AGN tori derived from
  CLUMPY torus models. Since these distributions combined over the
  total AGN population need to match the intrinsic type-2 AGN
  fraction, we reveal a population of X-ray undetected objects with
  high-covering factor tori, which are increasingly numerous at higher
  AGN luminosities. When these "missing" objects are included, we find
  that Compton-thick AGN account at most for 37$_{-10}^{+9}$\% of the
  total population. The intrinsic type-2 AGN fraction is 58$\pm$4\%
  and has a weak, non-significant (less than 2$\sigma$) luminosity
  dependence. This contradicts the results generally reported by AGN
  surveys, and the expectations from receding torus models. Our
  findings imply that the majority of luminous rapidly-accreting
  supermassive black holes at $z$$\leq$1 reside in highly-obscured
  nuclear environments but most of them are so deeply embedded that
  they have so far escaped detection in X-rays in $<$10 keV wide-area
  surveys.
\end{abstract}

\section{Introduction}\label{s:intro}
Dedicated searches for Active Galactic Nuclei (AGN) generally find
that the fraction of AGN classified either as optical type-2
(obscured) or X-ray absorbed, decreases substantially with increasing
luminosity~\citep[][]{lawrence82, hasinger05, simpson05, della08,
  burlon11, merloni14, ueda14, buchner15}.

To explain these findings, receding torus models have often been
adopted~\citep{lawrence91, simpson05, honig07}. They postulate that
the geometry of the material obscuring the AGN nuclear region, the
dusty torus~\citep{antonucci93, garcia16}, changes with AGN
luminosity. The torus geometrical covering factor (henceforth \ncv)
defines the fraction of the sky around the AGN central engine
that is obscured. If \cv decreases with increasing AGN luminosity,
then this considerably reduces the probability of finding luminous
type-2 AGN~\citep{elitzur12}.

  The observed decrease of the ratio of the torus infrared luminosity
  and the AGN bolometric luminosity (${\rm L_{torus}/L_{bol}}$) with
  ${\rm L_{bol}}$ has often been interpreted as direct evidence of a
  receding torus~(\citealt{maiolino07, treister08,lusso13}; but see
  \citealt{netzer16}). These results should be treated with caution
  since a one-to-one correspondence is not expected between ${\rm
    L_{torus}/L_{bol}}$ and \ncv~\citep{stalevski16}.

We can determine \cv using radiative transfer models that
self-consistently reproduce the emission from dust in the torus heated
by the
AGN~\citep[e.g.][]{fritz06,nenkova08,schartmann08,honig10,stalevski16}. Using
torus models with a clumpy distribution of dust
from~\citet[][henceforth N08]{nenkova08}, we determined, for the first
time, the probability density distributions of \cv for individual
objects~\citep[][henceforth M16]{mateos16} for a large, uniformly
selected, and complete flux-limited sample of X-ray selected AGN
drawn from the Bright Ultra-hard XMM-Newton
Survey~\citep[BUXS;][]{mateos12, mateos13}.

Using the distributions of \cv for the AGN in BUXS, we derive here the
intrinsic fraction of optical type-2 AGN at redshifts
0.05$\leq$$z$$\leq$1 as a function of intrinsic (absorption-corrected)
rest-frame 2-10 keV luminosity from ${\rm 10^{42}}$ to ${\rm
  10^{45}\,erg\,s^{-1}}$ (henceforth $L_{\rm X}$). We also investigate
whether the decrease of \cv with $L_{\rm X}$, which we observe in the
BUXS sample, is a property of the AGN population. Throughout, errors
are 1$\sigma$ (the 16th and 84th percentiles when referring to
distributions) unless otherwise stated. We adopt the concordance
cosmology, $\Omega_{\rm M}$=0.3, $\Omega_{\rm \lambda}$=0.7 and
$H_0$=70${\rm\,km\,s^{-1}\,Mpc^{-1}}$.

\section{AGN sample}\label{s:sample}
Our AGN sample is drawn from the BUXS survey. BUXS includes 255 X-ray
bright ($f_{4.5-10\,{\rm keV}}$$>$${\rm
  6\times10^{-14}\,erg\,cm^{-2}\,s^{-1}}$) AGN detected with
XMM-Newton in the 4.5-10 keV band over 44.43\,deg$^2$. Out of these,
252 have robust redshift $z$ and optical spectroscopic
  classifications. Objects with detected rest-frame UV/optical broad
emission lines (full width at half maximum $\geq$1500 km s${\rm
  ^{-1}}$) are classified as type-1 and those with narrower emission
lines as type-2.

Here, we only consider AGN with $L_{\rm X}$$\geq$${\rm
  10^{42}\,erg\,s^{-1}}$, to minimize host galaxy contamination,
by increasing the AGN to galaxy contrast ratio, and with $z$$\leq$1,
to avoid strong evolutionary effects. This restricts our sample to 199
objects, with ${\rm 10^{42}}$$\leq$$L_{\rm X}$$\leq$${\rm
  10^{45}\,erg\,s^{-1}}$ and 0.05$\leq$$z$$\leq$1.

BUXS is a unique survey to conduct this study. It is the only AGN
sample for which we know \cv for almost all ($\sim$99\%) objects. It
is sufficiently large to constrain accurately the intrinsic type-2 AGN
fraction. For all sources we have good quality X-ray spectroscopy
($\sim$few hundred counts) which guarantees robust estimates of
$L_{\rm X}$. Assuming the worst case, that all three unidentified
sources are in the $z$,$L_{\rm X}$ interval under study, they
represent at most $\sim$1.5\% of our sample. Clearly, our results are
not affected by identification incompleteness effects, that would bias
against obscured AGN.

To compute the luminosity dependence of the type-2 AGN fraction we
further divided our sample into three luminosity bins of equal
logarithmic width: ${\rm 10^{42}}$-${\rm 10^{43}}$, ${\rm
  10^{43}}$-${\rm 10^{44}}$ and ${\rm 10^{44}}$-${\rm
  10^{45}\,erg\,s^{-1}}$ (see Table~1).

\section{The covering factor of AGN tori}\label{f2}
In~\citet{mateos15} we built the rest-frame UV-to-infrared spectral
energy distributions (SED) of our objects using data from the Sloan
Digital Sky Survey~\citep[][]{abazajian09}, the Two Micron All Sky
Survey~\citep[][]{cutri03}, the UKIRT Infrared Deep Sky
Survey~\citep[][]{lawrence07}, and the Wide Field Infrared Survey
Explorer~\citep[WISE,][]{wright10}. With an SED decomposition analysis
we isolated the emission associated with dust in the torus heated by
the AGN at rest-frame wavelengths from $\sim$1\mic\, to $\sim$22\mic.

In M16 we fitted the torus SEDs with the N08 models using the Bayesian
inference tool BayesCLUMPY that provides posterior distributions for
all the free parameters of the models~\citep{asensio09}. We used
  truncated uniform prior distributions for all the torus model
  parameters (six in total) in the ranges listed in Table 1 from M16.

In the N08 models the dust is distributed in optically thick
  clouds ($\tau_V$$>$1 at 5500\,\AA). The torus inner radius is set
by the sublimation temperature of the dust grains ($\approx$1500
K). The radial distribution of clouds declines as a power-law. The
vertical angular distribution of clouds has no sharp boundary and it
is parameterized with a Gaussian. The geometrical covering factor of
the torus $f_2$ is defined as

\begin{eqnarray}
 {f_{\rm 2}=1-\int_0^{\pi/2} \! P_{\rm esc}(\beta)\,{\rm cos}(\beta){\it d}\beta}
\label{eq2}
\end{eqnarray}
where $P_{\rm esc}$ is the probability that light from the AGN will escape
without being absorbed at an angle $\beta$ from the torus equatorial
plane:
\begin{eqnarray}
 {P_{\rm esc}(\beta)=e^{-N_{\rm 0} \times e^{(-\beta^2/\sigma^2)}}.}
\label{eq1}
\end{eqnarray}
\cv depends on the angular width of the torus ($\sigma$) and the
  mean number of clouds along the equatorial direction ($N_0$). Using the
  posterior distributions of $\sigma$ and $N_0$ we calculated the
  probability density distribution of \cv for each source.

\section{Observed type-2 AGN fraction \lowercase{vs.} \ncv}\label{s:method}
Since \cv represents a geometrical covering factor, in any AGN
subpopulation having a given dust covering factor \ncv, the fraction
of type-2 objects should intrinsically be \ncv~\citep{elitzur12}.

We started by computing the observed type-2 AGN fraction in BUXS as a
function of \ncv, fully taking into account the uncertainties in
\ncv. We divided the range of \cv into five bins of width
$\Delta$\ncv=0.2, the 1$\sigma$ average error in our \cv estimates
from the individual distributions. For each source, we obtained the
fraction of \cv in each bin by integrating its \cv probability
distribution. The observed type-2 AGN fraction in bin $i$,
$F_{obs}^i$, is defined as,

\begin{equation}
  F_{obs}^i={\sum_{j=1}^{n_2}{F_{2,j}^i} \over {\sum_{j=1}^{n_2}{F_{2,j}^i} + \sum_{k=1}^{n_1}{F_{1,k}^i}}}
\end{equation}

\noindent
where $n_1$ and $n_2$ are the number of type-1 and type-2 AGN,
respectively (see Table~1) and $F_{1,k}^i$ and $F_{2,j}^i$ are the
fractions of the probability distributions of \cv in bin $i$.

  \begin{figure}[]
  \vspace{0.5cm}
\includegraphics[angle=-270,width=0.5\textwidth]{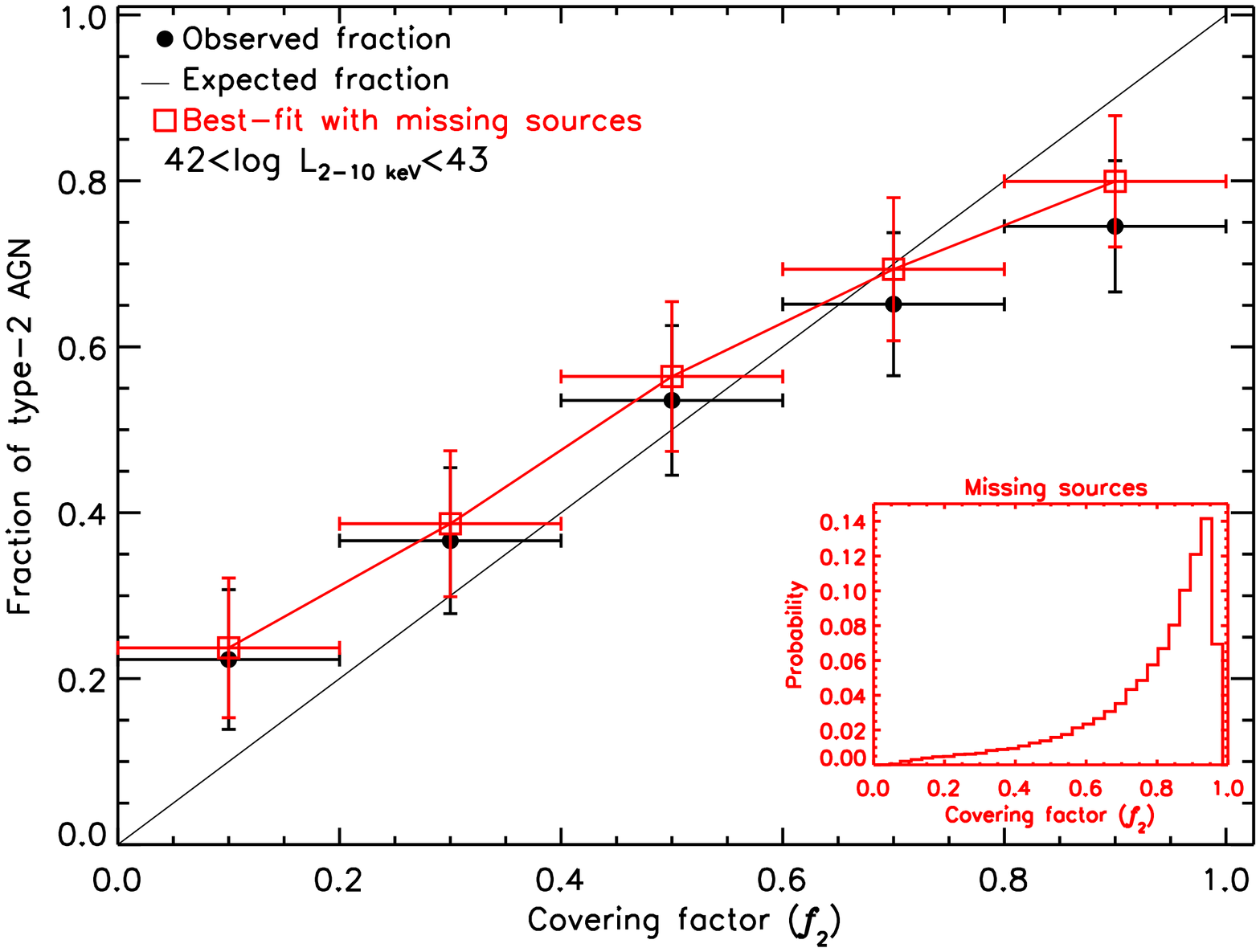}\\
\includegraphics[angle=-270,width=0.5\textwidth]{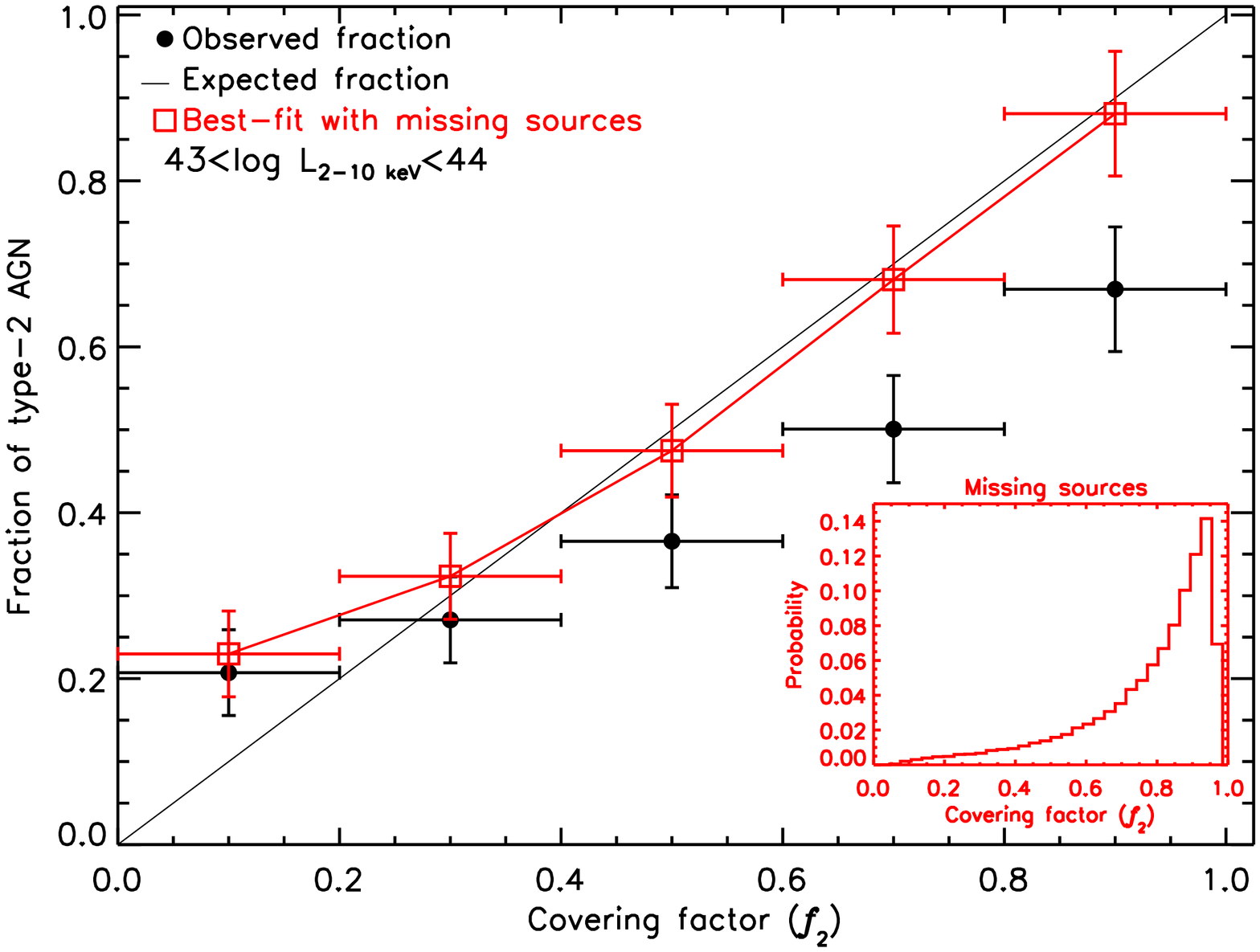}
\includegraphics[angle=-270,width=0.5\textwidth]{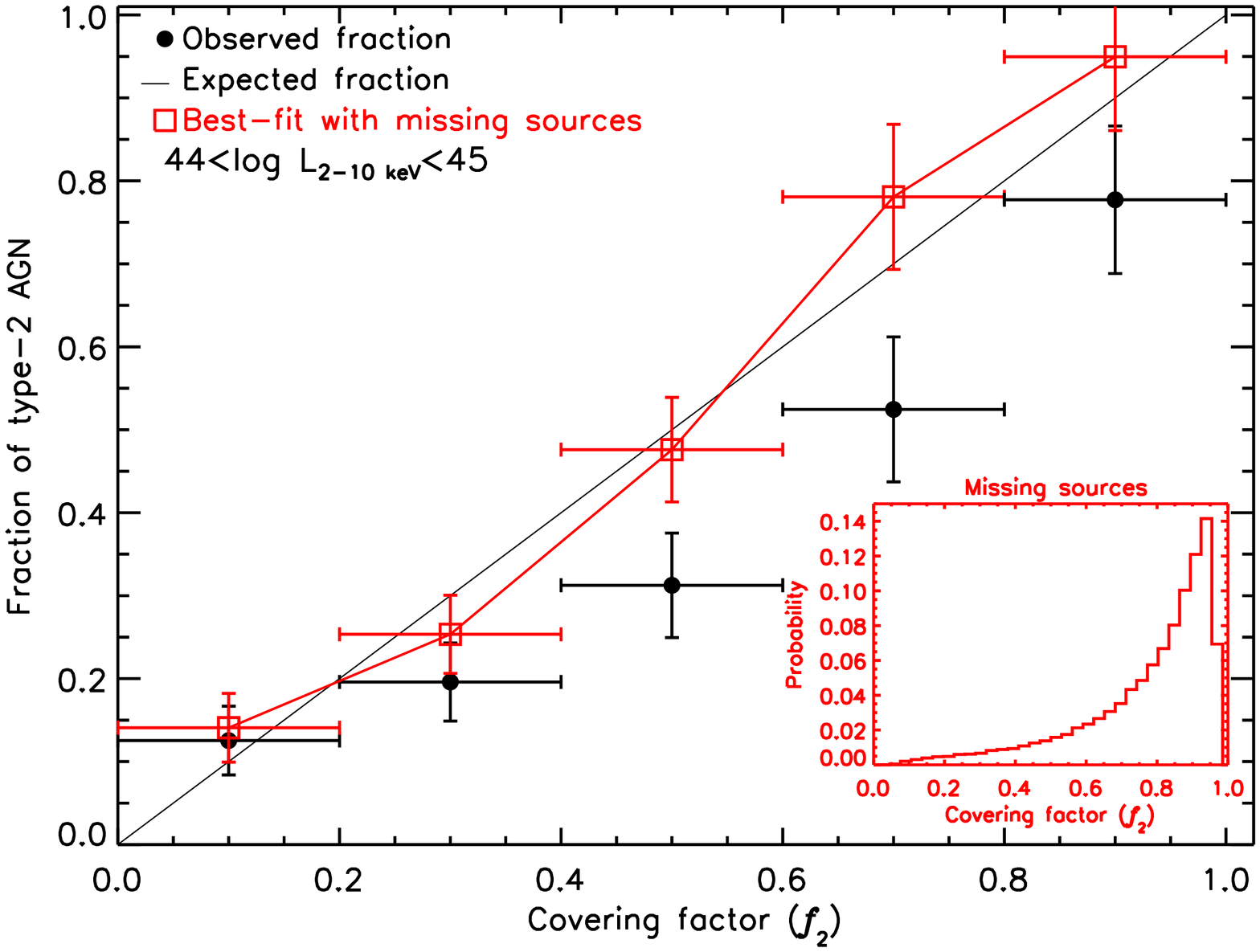}\\
\smallskip
\hspace{0.6cm}
\caption{Type-2 AGN fraction vs. torus covering factor $f_2$ for
  objects with ${\rm 10^{42}}$$<$$L{\rm _X}$$<$${\rm 10^{43}}$ (top),
  ${\rm 10^{43}}$$<$$L{\rm _X}$$<$${\rm 10^{44}}$ (middle) and ${\rm
    10^{44}}$$<$$L{\rm _X}$$<$${\rm 10^{45}\,erg\,s^{-1}}$
  (bottom). Filled circles are the observed type-2 AGN fractions in
  BUXS. Open squares are the best-fit models to the 1:1 relations
    (black solid lines) obtained by allowing a population of
    non-detected type-2 sources. The insets show the assumed \cv
    distribution of these missed sources.}
\vspace{0.5cm}
\label{fig:f1}
\end{figure}

To compute the uncertainties in $F_{obs}^i$, fully taking into account
both source Poisson counting noise and the uncertainties in \cv for
each source (the full \cv distributions), we used a bootstrap error
estimate. We generated ${\rm 10^6}$ mock samples by randomly selecting
type-1 and type-2 objects, with their corresponding \cv distributions,
from the original samples with replacement. Each mock sample
  contained a number of type-1 and type-2 AGN that was calculated from
  binomial distributions keeping constant the observed number of
  sources, i.e., ($n_1$+$n_2$), and assuming that the true type-2 AGN
  fraction is the observed one, $n_2$/($n_1$+$n_2$). We computed
$F_{obs}^i$ for each simulated dataset and then determined the median
and percentiles.

Since we will use a chi-squared ($\chi^{\rm 2}$) fit to derive the
intrinsic type-2 AGN fraction in the next section, we have corrected
for the small asymmetry in the $F_{obs}^i$ errors. For each set of
values ($F_{obs}$, $F_{obs}^{-}$, $F_{obs}^{+}$) we computed the
Gaussian function that has mean $F_{obs}$ and dispersion $\Delta
F_{obs}$ such that its integral from $F_{obs}$-$F_{obs}^{-}$ to
$F_{obs}$+$F_{obs}^{+}$ is 68.3\%. We used these $\Delta F_{obs}$ as
$\sigma$ errors in our $\chi^{\rm 2}$ fits.

Our results are illustrated in Fig.~\ref{fig:f1}. The y-axis
represents the observed fraction of AGN classified as optical type-2
in BUXS. The x-axis represents the covering factor of the torus
inferred from SED-fitting with N08 models. If BUXS did not miss any
highly-covered AGN (\ncv$\sim$1) our points should follow the 1:1
relation between the type-2 AGN fraction and \ncv. Clearly, this is
not the case, especially at $L{\rm _X}$$>$${\rm
  10^{43}\,erg\,s^{-1}}$. There are not enough luminous type-2 AGN
with high \ncv, therefore some must have escaped X-ray detection.

\section{Intrinsic fraction of type-2 AGN}\label{true_t2}
We have derived the global intrinsic type-2 AGN fraction by requiring
that the type-2 AGN fraction and \cv follow a 1:1 relation, i.e., for
each \cv the intrinsic fraction of type-2 AGN must be equal to
\ncv. To do so we made the following assumptions:

(i) The AGN missed are all type-2. BUXS is not biased against type-1
AGN\footnote[1]{Based on the Portable, Interactive, Multi-Mission
  Simulator (PIMMS) v4.8d assuming a power-law spectrum with photon
  index $\Gamma$=1.9, at the redshifts under study the decrease in the
  4.5-10 keV count-rate from $N_{\rm H}$=0 to X-ray absorbing column
  densities of $N_{\rm H}$=${\rm 10^{22}\,cm^{-2}}$ and $N_{\rm
    H}$=${\rm 10^{23}\,cm^{-2}}$ is $<$1\% and $\lesssim$3\%,
  respectively.}. Moreover, the AGN missed cannot be "X-ray weak" AGN,
not only because these are a rare population~\citep[e.g.][]{brandt00,
  risaliti03}, but also because there is no physical reason why such
objects should have tori with the highest \cv among all
AGN. Flux-limited surveys below 10 keV are incomplete for low-$z$ AGN
whose line of sight X-ray absorption approaches the Compton-thick
limit ($N_{\rm H}$=${\rm 1.5\times10^{24}\,cm^{-2}}$). Thus, the
type-2 AGN missed are either Compton-thick or heavily-absorbed ($N_{\rm
  H}{\rm \sim a\,few \times10^{23}\,cm^{-2}}$).

(ii) The stacked probability distribution of \cv for all the
type-2 AGN in BUXS with $N_{\rm H}$$>$$4\times10^{23}\,{\rm cm^{-2}}$
(insets in Fig.~\ref{fig:f1}) represents well \cv in the objects
missed. This is well-justified since heavily-absorbed type-2 AGN have
\ncv$\sim$1~\citetext{\citealt{ramos11}; M16} and their distributions
of \cv are all very similar, regardless of $L_{\rm
  X}$~\citetext{\citealt{alonso11,ichikawa15}; M16}. If we assume
instead a distribution of \cv peaking at smaller values, our main
results remain unchanged, although only poorer fits are possible.

\begin{table*}[t]
\begin{center}
  \caption{Summary of the properties of our AGN samples and the results of our analysis.}
  \begin{tabular}{cccccccccccc}
\hline
  \footnotesize log($L_{\rm X}$) &  \footnotesize $n_1$ & \footnotesize $n_2$ & \footnotesize $\langle$$L_{\rm X}$$\rangle_{\rm 1}$  & \footnotesize $\langle$$z$$\rangle_{\rm 1}$ & \footnotesize $\langle$$L_{\rm X}$$\rangle_{\rm 2}$ & \footnotesize $\langle$$z$$\rangle_{\rm 2}$ & \footnotesize Observed & \footnotesize $N_2$ & \footnotesize Intrinsic & \footnotesize Type-2 fraction & \footnotesize Compton-thick\\
  \footnotesize  &  \footnotesize  & \footnotesize  & \footnotesize  & \footnotesize & \footnotesize  & \footnotesize  & \footnotesize  type-2 fraction & \footnotesize & \footnotesize type-2 fraction & \footnotesize missed & \footnotesize fraction\\
  \footnotesize (1) &  \footnotesize (2) & \footnotesize (3) & \footnotesize (4)  & \footnotesize (5) & \footnotesize (6)  & \footnotesize (7) & \footnotesize (8) & \footnotesize (9) & \footnotesize (10) & \footnotesize (11) & \footnotesize (12)\\
  \hline
  42-43 & 16 & 21 & 42.75 & 0.10 & 42.80 & 0.11 & 56.8$_{-8.4}^{+7.5}$ &  5$_{-4}^{+6}$   & 64.4$_{-9.3}^{+7.1}$   &  25.3$_{-14.1}^{+16.7}$  & $\leq$14.5$_{-7.7}^{+14.6}$\\
  43-44 & 53 & 33 & 43.59 & 0.29 & 43.48 & 0.25 & 38.4$_{-5.0}^{+5.4}$ & 38$_{-9}^{+18}$  & 58.7$_{-6.8}^{+5.6}$   &  56.9$_{-12.2}^{+7.5}$   & $\leq$32.6$_{-8.3}^{+8.1}$\\
  44-45 & 55 & 21 & 44.50 & 0.76 & 44.41 & 0.62 & 27.6$_{-4.5}^{+5.6}$ & 40$_{-14}^{+19}$ & 54.6$_{-8.7}^{+6.8}$   &  69.7$_{-14.4}^{+6.1}$   & $\leq$37.0$_{-10.5}^{+8.9}$\\
\hline 
\end{tabular}
\end{center}
Notes.---Column 1: X-ray luminosity range in ${\rm erg\,s^{-1}}$ in
logarithmic units; columns 2 and 3: number of type-1 and type-2
AGN in the bin, respectively; columns 4 to 7: median X-ray luminosity
and redshift of type-1 and type-2 AGN, respectively; column 8:
observed type-2 AGN fraction; column 9: number of type-2 AGN missed;
column 10: intrinsic type-2 AGN fraction; column 11: fraction of
type-2 AGN that have escaped X-ray detection; column 12: fraction of
type-2 AGN that have escaped X-ray detection over the total
population. Fractions are given in percentage units.
\label{tab:bins}
\smallskip
\end{table*}
\smallskip

To compute the number of type-2 AGN missed, $N_2$, we increased $N_2$
until we found the best $\chi^{\rm 2}$ fit to the 1:1 relations in
Fig.~\ref{fig:f1}. The best-fit type-2 AGN fractions, $F_{bf}^i$ (open
squares), are

\begin{equation}
  F_{bf}^i={{\sum_{j=1}^{n_2}{F_{2,j}^i} + {F_{CT}^i}\times N_2} \over {\sum_{j=1}^{n_2}{F_{2,j}^i} + \sum_{k=1}^{n_1}{F_{1,k}^i} + {F_{CT}^i}\times N_2}}
\end{equation}

\noindent
where $F_{CT}^i$ is the fraction of the \cv distribution used to
represent the missing AGN in the bin $i$. We obtained the
uncertainties on $N_2$ from our fits as $P(N_2)\propto
e^{-{\Delta\chi^2}/2}$. Next, we computed the intrinsic type-2 AGN
fraction, $F_{intr}$=($n_2$+$N_2$)/($n_1$+$n_2$+$N_2$), with
uncertainties using a Bayesian approach~\citep{wall03}. We have
assumed a binomial distribution similar to that in
Section~\ref{s:method} weighted with $P(N_2)$. Since $P(N_2)$ are only
defined for $N_{\rm 2}$$\ge$0, our estimates of $F_{intr}$ are higher
than what would be directly obtained from the numbers listed in
columns 2, 3 and 9 of Table~1. We have followed this same Bayesian
approach to compute the fractions listed in columns 8, 11 and 12 in
Table~1.

Since the errors of $F_{obs}^i$ are not independent we tested the
robustness of our $\chi^{\rm 2}$ fitting with Monte Carlo
simulations. We used our simulated values of $F_{obs}^{i}$ (see
Sec.~\ref{s:method}) to determine $N_2$ with a least squares fitting,
and found that the results were indistinguishable from those obtained
with the $\chi^{\rm 2}$ fitting.

  We now discuss some issues that could affect our results.

In our computations of $F_{obs}^i$, objects with \ncv$\sim$0.5
contribute to both the upper and lower \cv neighboring bins while
objects with extreme \cv values contribute only towards the central
bins. This effect could flatten our $F_{obs}^{i}$ estimates in
Fig.~\ref{fig:f1} with respect to the 1:1 relation removing the need
of adding any missing sources. To address this issue, we have used
Monte Carlo simulations of source samples of the same size as ours but
following the 1:1 relation. The simulated objects are assigned
  \cv probability distributions drawn from the real type-1/type-2
  objects, proportionally to their weight at the needed \cv
  value. This results only in a difference of a few percent in our
  $F_{intr}$ estimates, not changing our conclusions.

For two type-1 AGN with $L_{\rm X}$$>$$10^{43}{\rm \,erg\,s^{-1}}$ we
could not determine \cv because they were not detected by WISE. We
included both sources in our analysis by assigning them \cv
distributions drawn at random with replacement from the type-1 AGN
sample in the corresponding luminosity bin. Since all type-1 AGN have
similar distributions of \cv (M16) our approach is well
justified. Therefore, our results are free from systematic
uncertainties associated with the lack of detection of X-ray sources
at infrared wavelengths.

Type-2 AGN typically have optical extinctions associated with dust in
their hosts of $A_V$$\leq$5
mag~\citep[e.g.][]{alonso03,alonso11}. These are too small to have any
noticeable effect on our torus SEDs and thus on our \cv
estimates. Host galaxy dilution may cause a type-1 AGN to be
misclassified as type-2, especially at low
luminosities~\citep{caccianiga07}. If there are cases such as these in
BUXS, the number is too small to affect our results (M16).

By restricting our analysis to AGN with $L_{\rm X}$,$z$ where BUXS is
complete, we found that the uncertainties were larger, but the results
remained unchanged. Finally, we have verified that if we force type-1
and type-2 AGN to have the same $z$ distributions on each $L{\rm _X}$
bin (with a bootstrap re-sampling), the results remain the same. Thus,
if type-1 and type-2 AGN evolve differently at the $z$ under
study~\citep[e.g.][]{reyes08}, this has no noticeable impact on
our analysis.

\begin{figure*}[t]
  \centering
  \begin{tabular}{cc}
    \hspace{-0.7cm}\includegraphics[angle=-270,width=0.7\textwidth]{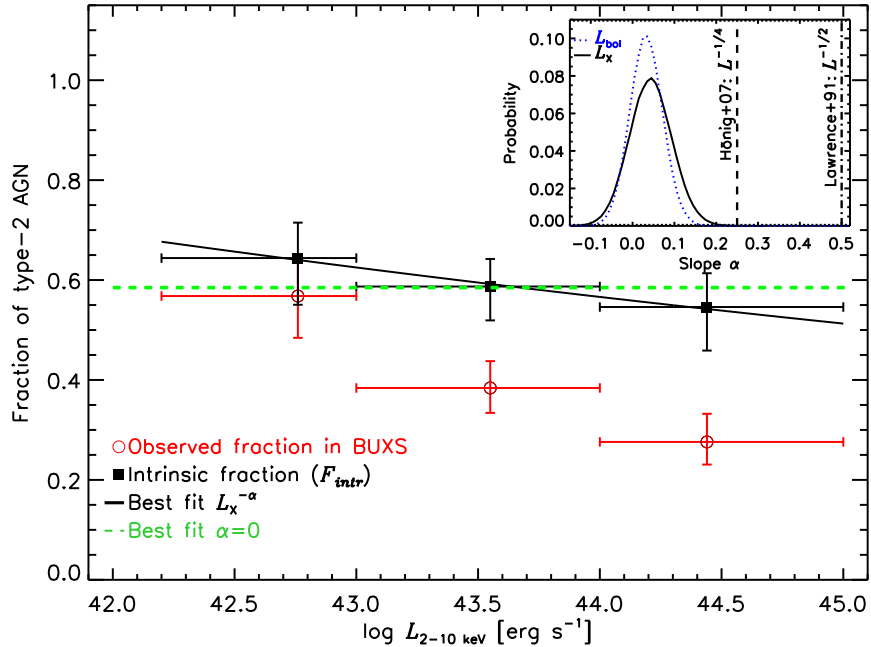}
  \end{tabular}
  \vspace{-0.3cm} \caption {Luminosity dependence of the observed
    (circles) and intrinsic (squares) type-2 AGN fractions. The dashed
    line is the best fit to the intrinsic type-2 AGN fraction assuming
    no luminosity dependence. The solid line is the best fit assuming
    a power-law dependence. Inset: probability density distributions of
    the best-fit power-law index, $\alpha$. The vertical lines
    represent the values of $\alpha$ expected in the receding torus
    models of~\citet[][$\alpha$$\sim$0.5]{lawrence91}
    and~\citet[][$\alpha$$\sim$0.25]{honig07}.}
  \label{fig:f2}
\end{figure*}

\section{Compton-thick AGN fraction}\label{f2_CT}
A non-negligible fraction of luminous type-2 AGN with high $f_2$ have
escaped X-ray detection. The fraction of AGN missed over the total
population (last column in Table~1) gives us a strict upper limit to
the Compton-thick fraction, since none of the AGN in BUXS are
Compton-thick. At $L{\rm_X}$$>$${\rm 10^{43}\,erg\,s^{-1}}$
Compton-thick AGN cannot contribute more than $37.0_{-10.5}^{+8.9}$\%
to the total AGN population, in agreement with recent
estimates~\citep{buchner15, ricci15}. At $L{\rm_X}$$<$${\rm
  10^{43}\,erg\,s^{-1}}$ ($z$$\sim$0.1) our results are also
consistent, within the uncertainties, with Compton-thick fractions
reported for low-$z$ AGN
samples~\citep{bassani06,burlon11,ricci15,akylas16}.

\section{Luminosity dependence of the type-2 AGN fraction}\label{f2_l}
Fig.~\ref{fig:f2} shows the luminosity dependence of $F_{intr}$. For
comparison, we also show the observed type-2 AGN fraction in BUXS
which, as typically found in flux-limited X-ray surveys, decreases
substantially with $L{\rm _X}$. We have parameterized the luminosity
dependence of $F_{intr}$ with a power-law of the form
$F_{intr}$\,$\propto$\,$L_{\rm X}^{-\alpha}$. A simple $\chi^2$ fit to
a straight line in log-log space using the values and errors of
$F_{intr}$ in Table~1 yields the relation,

\begin{equation}
  {\rm log}\,F_{intr}=-(0.043_{-0.051}^{+0.051})\times {\rm log}\,L_{\rm X}+(1.645_{-2.218}^{+2.235})
\end{equation}
\noindent

The inset in Fig.~\ref{fig:f2} shows the probability density
distribution of $\alpha$ which we find to be relatively narrow around
a peak close to zero (3$\sigma$ upper limit $\alpha$$<$0.19). When the
missing objects are included, the dependence of $F_{intr}$ on $L{\rm
  _X}$ is no longer significant (just 80.1\% of our simulations gave
$\alpha$$>$0). A fit with constant $F_{intr}$ yields instead
$F_{intr}$=${\rm 58.5_{-4.4}^{+4.1}}$\% (see Fig.~\ref{fig:f2}).

Since most of the radiation absorbed by the dust in the torus is
emitted by the accretion disk at UV/optical wavelengths, to test the
validity of receding torus models we should investigate whether
$F_{intr}$ varies with AGN bolometric power, $L_{\rm bol}$. If we
assume a constant X-ray bolometric correction, our results remain
unchanged. If we adopt instead that the conversion from $L{\rm_X}$ to
$L_{\rm bol}$ increases with luminosity~\citep{marconi04, hopkins07},
using one such bolometric correction~\citep{marconi04}, our AGN span
almost four orders of magnitude in $L_{\rm bol}$ (from ${\rm 10^{43}}$
to 7$\times$${\rm 10^{46}\,erg\,s^{-1}}$) yielding an even tighter
limit on the slope, $\alpha$=$0.033_{-0.039}^{+0.040}$ (3$\sigma$
upper limit $\alpha$$<$0.14). We find evidence that $F_{intr}$ does
not decrease with increasing luminosity, clearly contradicting the
expectations of receding torus models.

In our analysis we have not considered that the uncertainties in
$L{\rm _X}$ might move AGN between luminosity bins. Because the
bins are 1 dex wide and the uncertainties in $L{\rm_X}$ small (medians
of 5\% and 35\% for type-1 and type-2 AGN, respectively), the effect
should be negligible. A simple calculation of the scatter of sources
in the bins by counting sources weighted by the distributions of
$L{\rm _X}$ revealed that, at $L{\rm _X}$$>$${\rm
  10^{43}\,erg\,s^{-1}}$ our results do not change. At $L{\rm
    _X}$$<$${\rm 10^{43}\,erg\,s^{-1}}$ $N_2$ increases to $\sim$nine
  which translates into a $F_{intr}$ of $\sim$67\%. This has a
  negligible impact on our estimate of $\alpha$
  ($\Delta$$\alpha$$\sim$0.01).

By filling the gap between studies in the local Universe and at cosmic
epochs when supermassive black hole (SMBH) mass growth peaked,
$z$$\sim$1-2~\citep{martinez05, ueda14, aird15, assef15, buchner15,
  delmoro16}, we show here that luminous highly-obscured AGN dominate
the population of fast growing SMBH up to $z$$\sim$2. Surveys with the
Nuclear Spectroscopic Telescope Array (NuSTAR) have started to reveal
this elusive AGN population. At the $z$ of our sample and luminosities
$2\times10^{43}$$<$$L_{\rm 10-40\,keV}$$<$${\rm
  2\times10^{44}\,erg\,s^{-1}}$ the observed type-2 AGN fraction is
consistent with our findings~\citep[$53_{-15}^{+14}$\% assuming
  $L_{\rm 10-40\,keV}$/$L_{\rm 2-10\,keV}$$\approx$1;][]{lansbury16}.

\section{Summary and conclusions}
We have determined the intrinsic type-2 AGN fraction at
0.05$\leq$$z$$\leq$1 and at ${\rm 10^{42}}$$\leq$$L_{\rm
  2-10\,keV}$$\leq$${\rm 10^{45}\,erg\,s^{-1}}$. To do so we used a
complete flux-limited sample of 199 X-ray selected AGN drawn from the
Bright Ultra-hard XMM-Newton Survey (BUXS). For this sample we have
robust estimates of the geometrical covering factor of the torus in
the framework of N08 clumpy torus models.

Since the distribution of covering factors needs to match the fraction
of optical type-2 AGN, we reveal the existence of a substantial
population of X-ray undetected objects with high-covering factor tori,
which are increasingly numerous at higher AGN luminosities. When
these "missing" objects are included, Compton-thick AGN account at
most for 37.0$_{-10.5}^{+8.9}$\% of the total AGN population, in
agreement with previous estimates. We find that the intrinsic type-2
AGN fraction is $58.5_{-4.4}^{+4.1}$\% and has a weak and
non-significant (less than 2$\sigma$) luminosity dependence. This is
in clear contradiction with the results generally reported by AGN
surveys and the expectations from receding torus models.

Our findings imply that the majority of luminous, rapidly-accreting
SMBH, reside in highly-obscured nuclear environments, many so deeply
buried that they remain undetected in X-rays at the depths of $<$10
keV wide-area surveys.

\smallskip
We thank the anonymous referee for the complete and deep revision of
our manuscript, that helped us to substantially improve our work. SM
acknowledges financial support through grant
AYA2016-76730-P. (MINECO/FEDER). FJC, XB and A.A.- H. acknowledge
financial support through grant AYA2015-64346-C2-1-P (MINECO/FEDER).
AH-C acknowledges financial support through grants AYA2015-70815-ERC
and AYA2012-31277. CRA acknowledges financial support through grant
AYA2016-76682-C3-2-P and the Ram\'on y Cajal Program through project
RYC-2014-15779 (MINECO). TM is supported by CONACyT Grants 179662,
252531 and UNAM-DGAPA PAPIIT IN104216. Based on observations collected
at the European Organization for Astronomical Research in the Southern
hemisphere, Chile. Based on observations made with the William
Herschel Telescope-operated by the Isaac Newton Group, the Telescopio
Nazionale Galileo-operated by the Centro Galileo Galilei and the Gran
Telescopio de Canarias installed in the Spanish Observatorio del Roque
de los Muchachos of the Instituto de Astrof\'isica de Canarias.


\begin{thebibliography}{}

\bibitem[Abazajian et al.(2009)]{abazajian09} Abazajian, K. N.,
  Adelman-McCarthy, J. K., Ag\"{u}eros, M. A. et al. 2009, \apjs, 182, 543

\bibitem[Aird et al.(2015)]{aird15} Aird, J., Alexander, D. M.,
  Ballantyne, D. R., Civano, F. et al. 2015, \apj, 815, 66

\bibitem[Akylas et al.(2016)]{akylas16} Akylas, A., Georgantopoulos,
  I., Ranalli, P. et al. 2016, \aap, 594, 73

\bibitem[Alonso-Herrero et al.(2003)]{alonso03} Alonso-Herrero, A., Quillen,
  A. C., Rieke, G. H., Ivanov, V. D., Efstathiou, A. 2003,
  \aj, 126, 81
  
\bibitem[Alonso-Herrero et al.(2011)]{alonso11} Alonso-Herrero, A., Ramos
  Almeida, C., Mason, R., Asensio Ramos, A. et al. 2011,
  \apj, 736, 82

\bibitem[Antonucci(1993)]{antonucci93} Antonucci, R. 1993, \araa, 31,
  473

\bibitem[Asensio Ramos \& Ramos Almeida(2009)]{asensio09} Asensio
  Ramos, A., \& Ramos Almeida, C. 2009, \apj, 696, 2075
  
\bibitem[Assef et al.(2015)]{assef15} Assef, R.~J., Eisenhardt,
  P.~R.~M., Stern, D. et al. 2015, \apj, 804, 27A
  
\bibitem[Bassani et al.(2006)]{bassani06} Bassani, L., Molina, M.,
  Malizia, A. et al. 2006, \apj, 636, L65-L68

\bibitem[Brandt, Laor \& Wills (2000)]{brandt00} Brandt, W. N., Laor,
  A., Wills, B.J. 2000, \apj, 528, 637

\bibitem[Buchner et al.(2015)]{buchner15} Buchner, J., Georgakakis,
  A., Nandra, K. et al. 2015, \apj, 802, 89

\bibitem[Burlon et al.(2011)]{burlon11} Burlon, D., Ajello, M.,
  Greiner, J., Comastri, A., Merloni, A., Gehrels, N. 2011, \apj,
  728, 58

\bibitem[Caccianiga et al.(2007)]{caccianiga07} Caccianiga, A.,
  Severgnini, P., Della Ceca, R., Maccacaro, T. et al. 2007, \aap,
  470, 557

\bibitem[Cutri et al.(2003)]{cutri03} Cutri, R. M., Skrutskie, M. F.,
  van Dyk, S. et al. 2003, The IRSA 2MASS All-Sky Point Source
  Catalog, NASA/IPAC Infrared Science Archive,
  http://irsa.ipac.caltech.edu/applications/Gator/
  
\bibitem[Della Ceca et al.(2008)]{della08} Della Ceca, R., Caccianiga,
  A., Severgnini, P. et al. 2008, \aap, 487, 119

\bibitem[Del Moro et al.(2016)]{delmoro16} Del Moro, A.,
  Alexander, D. M., Bauer, F. E. et al. 2016, \mnras, 456, 2105

\bibitem[Elitzur(2012)]{elitzur12} Elitzur, M. 2012, \apj, 747,
  33
  
\bibitem[Fritz et al.(2006)]{fritz06} Fritz, J., Franceschini A.,
  Hatziminaoglou E. 2006, \mnras, 366, 767
  
\bibitem[Garc\'ia-Burillo et al.(2016)]{garcia16}
  Garc\'ia-Burillo, S., Combes, F., Ramos Almeida, C. et al. 2016,
  \apjl, 823, L12

\bibitem[Hasinger et al.(2005)]{hasinger05} Hasinger, G., Miyaji, T. \&
  Schmidt, M. 2005, \aap, 441, 417

\bibitem[H\"onig et al.(2007)]{honig07} H\"onig, S. F. \&
  Beckert, T.  2007, \mnras, 380, 1172

\bibitem[H\"onig \& Kishimoto(2010)]{honig10} H\"onig, S. F. \&
  Kishimoto, M. 2007, \aap, 523, 27

\bibitem[Hopkins et al.(2007)]{hopkins07} Hopkins,
  P.~F.,Richards, G.~T., Hernquist, L. 2007, \apj, 654, 731

\bibitem[Ichikawa et al.(2015)]{ichikawa15} Ichikawa, K., Packham, C.,
  Ramos Almeida, C., Asensio Ramos, A. et al. 2015, \apj, 803, 571
  
\bibitem[Lansbury et al.(2016)]{lansbury16} Lansbury, G. B., Stern,
  D., Aird, J., Alexander, D. M. et al. 2017, \apj, 836, 99

\bibitem[Lawrence \& Elvis(1982)]{lawrence82} Lawrence, A. \& Elvis,
  M. 1982, \apj, 256, 410
  
\bibitem[Lawrence et al.(1991)]{lawrence91} Lawrence, A. 1991, \mnras,
  252, 586

\bibitem[Lawrence et al.(2007)]{lawrence07} Lawrence, A., Warren,
  S. J., Almaini, O. et al. 2007, \mnras, 379, 1599


\bibitem[Lusso et al.(2013)]{lusso13} Lusso, E. \& Hennawi, J. F. \&
  Comastri, A. et al. 2013, \apj, 777, 86

\bibitem[Maiolino et al.(2007)]{maiolino07} Maiolino, R., Shemmer, O.,
  Imanishi, M. et al. 2007, \aap, 468, 979

\bibitem[Marconi et al.(2004)]{marconi04} Marconi, A., Risaliti,
  G., Gilli, R. et al. 2004, \mnras, 351, 169

\bibitem[Martinez-Sansigre et al.(2005)]{martinez05}
  Martinez-Sansigre, A., Rawlings, S., Lacy, M. et al. 2005, \nat,
  436, 666

\bibitem[Mateos et al.(2012)]{mateos12} Mateos, S., Alonso-Herrero,
  A., Carrera, F. J. et al. 2012, \mnras, 426, 327

\bibitem[Mateos et al.(2013)]{mateos13} Mateos, S., Alonso-Herrero,
  A., Carrera, F. J. et al. 2013, \mnras, 434, 941

\bibitem[Mateos et al.(2015)]{mateos15} Mateos, S., Carrera, F. J.,
  Alonso-Herrero, A. et al. 2015, \mnras, 449, 1422
 
\bibitem[Mateos et al.(2016)]{mateos16} Mateos, S. , Carrera, F. J.,
  Alonso-Herrero, A. et al. 2016, \apj, 819, 166

\bibitem[Merloni et al.(2014)]{merloni14} Merloni, A., Bongiorno,
  A., Brusa, M. et al. 2014, \mnras, 437, 3550

\bibitem[Nenkova et al.(2008)]{nenkova08} Nenkova, M., Sirocky, M. M.,
  Nikutta, R., Ivezi\'c, Z., \& Elitzur, M. 2008, \apj, 685, 160
  
\bibitem[Netzer et al.(2016)]{netzer16} Netzer, H., Lani, C.,
  Nordon, R. et al. 2016, \apj, 819, 123

\bibitem[Ramos Almeida et al.(2011)]{ramos11} Ramos Almeida,
  C., Levenson, N. A., Alonso-Herrero, A. et al. 2011, ApJ, 731, 92

\bibitem[Reyes et al.(2008)]{reyes08} Reyes, R., Zakamska, N. L.,
  Strauss, M. A., Green, J. et al. 2008, \aj, 136, 2373

\bibitem[Ricci et al.(2015)]{ricci15} Ricci, C., Ueda, Y., Koss,
  M. J., Trakhtenbrot, B.  et al. 2015, \apj, 815, 13

\bibitem[Risaliti et al.(2003)]{risaliti03} Risaliti, G., Elvis, M.,
  Gilli, R., Salvati, M. 2003, \apj, 587, L9-13
  

\bibitem[Schartmann et al.(2008)]{schartmann08} Schartmann, M.,
  Meisenheimer, K., Camenzind, M., Wolf, S. et el. 2008, \aap, 482, 67

\bibitem[Simpson(2005)]{simpson05} Simpson, C. \mnras, 360, 565 

\bibitem[Stalevski et al.(2016)]{stalevski16} Stalevski, M.,
  Ricci, C., Ueda, Y. et al. 2016, \mnras, 458, 2288

\bibitem[Treister, Krolik \& Dullemond(2008)]{treister08} 
  Treister, E., Krolik, J. H. \& Dullemond, C. 2008, \apj, 679, 140

\bibitem[Ueda et al.(2014)]{ueda14} Ueda, Y., Akiyama, M.,
  Hasinger, G., Miyaji, T. \& Watson, M. G. 2014, \apj, 786, 104

\bibitem[Wall \& Jenkins(2003)]{wall03} Wall, J. V. \& Jenkins,
  C. R. 2003, Practical Statistics for Astronomers, Cambridge,
  Cambridge Univ. Press

\bibitem[Wright et al.(2010)]{wright10} Wright, E. L.,
  Eisenhardt, P. R. M., Mainzer, A. K. et al. 2010, \aj, 140, 1868

\end{thebibliography}
\end{document}